\documentclass[a4paper,11pt]{article}
\pdfoutput=1 % if your are submitting a pdflatex (i.e. if you have
\usepackage{jcappub} 
\usepackage[T1]{fontenc} 

\usepackage{natbib}
\usepackage{cancel}
\usepackage{enumitem}    
\usepackage{amsthm}         
\usepackage{amsmath} 	   
\usepackage{amssymb}       
\usepackage{amsfonts}
\usepackage{graphicx} 
\usepackage{dblfloatfix}	    
\usepackage{verbatim}
\usepackage{epstopdf}
\usepackage{tensor}
\usepackage{epsfig,multicol,bbm}
\usepackage{color}
\usepackage{colortbl}
\usepackage{float}
\usepackage{multirow}
\usepackage{array}
\usepackage{xcolor,colortbl}

\usepackage{subfig}
\usepackage{subcaption}

\usepackage{afterpage}

\allowdisplaybreaks

\title{Smoothing and flattening the universe through slow contraction versus inflation}
\author[a]{Anna Ijjas,}  
\author[*\,b,c]{Paul J. Steinhardt,}
\note[*]{Corresponding author: steinh@princeton.edu}
\author[d]{David Garfinkle}
\author[e]{and William G. Cook}

\affiliation[a]{Center for Cosmology and Particle Physics, Department of Physics, New York University, New York, NY 10003, USA}
\affiliation[b]{Department of Physics, Princeton University, Princeton, NJ 08544, USA}
\affiliation[c] {Jefferson Physical Laboratory, Harvard University, Cambridge MA 02138 USA}
\affiliation[d]{Department of Physics, Oakland University, Rochester, MI 48309, USA}
\affiliation[e]{Theoretisch-Physikalisches Institut, Friedrich-Schiller-Universit\"at Jena, 07743, Jena, Germany}

\emailAdd{ijjas@nyu.edu, steinh@princeton.edu, garfinkl@oalkland.edu,william.cook@uni-jena.de}

\abstract{  In a systematic study, we use an equivalent pair of improved numerical relativity codes based on a tetrad-formulation of the classical Einstein-scalar field equations to examine whether slow contraction or inflation (or both) can resolve the homogeneity, isotropy and flatness problems.  Our finding, based on a set of gauge/frame invariant diagnostics, is that slow contraction robustly and rapidly smooths and flattens spacetime  
beginning from initial conditions that are outside the perturbative regime of the flat Friedmann-Robertson-Walker  metric, whereas inflation fails these tests.  We present new numerical evidence supporting the conjecture that  the combination of  ultralocal evolution and an effective equation-of-state with pressure much greater than energy density is the key  to having robust and rapid smoothing.  The opposite of ultralocality occurs in expanding spacetimes, which is the  leading obstruction to smoothing following a big bang.    
}


\begin{document}
\maketitle 
\raggedbottom

\section{Introduction}
\label{sec1}

The foremost challenge for any theory of the origin and early evolution of the universe is to explain its homogeneity, isotropy and spatial flatness on large scales.  These features are extraordinary for a theory based on general relativity in which a foundational notion is that spacetime is malleable, capable of curving, shearing, fluctuating and warping.  Yet all  the successful predictions of $\Lambda$CDM (a.k.a. the basic “big bang theory”) – primordial nucleosynthesis, the cosmic microwave background, structure formation, and Hubble expansion –  rely on the assumption that there was no sign of this malleability by the time primordial nucleosynthesis commenced.   Instead, spacetime was well-described by a surprisingly simple flat Friedmann-Robertson-Walker (FRW) metric that is highly non-generic in general relativity.

In this paper, we present the results of a systematic comparative study of two mechanisms proposed to smooth and flatten the early universe beginning from initial conditions that are outside the perturbative regime of the flat FRW metric, namely inflation and slow contraction.  Inflation is a period of accelerated expansion hypothesized to occur shortly after the big bang during which the equation of state $\varepsilon \equiv (3/2)(1+ p/\rho) <1$ (where $p$ is the pressure and $\rho$ is the energy density of the dominant energy component) and the FRW scale factor $a(\tau) \propto \tau^{1/\varepsilon}$ grows faster with $\tau$ than the Hubble radius $H^{-1}(\tau) \propto \tau$.   Slow contraction is a period in which $a(\tau) \propto |\tau|^{1/\varepsilon}$ where $\varepsilon \gg 3$, which shrinks more slowly than the Hubble radius $|H^{-1}| \propto |\tau|$ as $\tau\rightarrow 0^{-}$. (We follow the convention of defining $\tau$ as positive and increasing during expansion and as negative and increasing towards $\tau \rightarrow 0^{-}$ during contraction. This is the reason for the absolute value signs in the expressions for contraction.)  
In either case, the requisite value of $\varepsilon$ can be attained by a scalar field $\phi$ evolving down a potential $V(\phi)$, for which $\varepsilon \rightarrow \frac{3}{2} \dot{\phi}^2/(\frac{1}{2}  \dot{\phi}^2 + V)$ in the homogeneous limit.  For inflation, obtaining $\varepsilon <1$ requires a positive potential energy and small scalar field kinetic energy;  for slow contraction, the potential energy must be negative and the kinetic energy must be large enough that the total energy density is positive.   Note that both require similar ingredients (Einstein gravity and a scalar field with a potential) and  the same number of initial conditions to be specified at the start, which makes for a fair comparison.  

For both proposed smoothing mechanisms, it is straightforward to show that they can smooth out small perturbations to a pure FRW metric, which is why both appeared at first to be promising candidates.   The true test, though, is to begin far from FRW conditions and solve the full set of non-linearly coupled partial differential Einstein-scalar equations of motion to determine if smoothing and flattening occurs.  This has become possible by adapting the numerical relativity tools originally introduced to study black hole inspiral and merger to address the  homogeneity and isotropy problem in early universe cosmology.   There have been a number of pioneering studies of inflation \cite{East:2015ggf,Clough:2016ymm,Aurrekoetxea:2019fhr,Clough:2017efm,Corman:2022alv,Garfinkle:2023vzf} and slow contraction \cite{Garfinkle:2008ei,Cook:2020oaj,Ijjas:2020dws,Ijjas:2021gkf,Ijjas:2021wml,Ijjas:2021zyf,Kist:2022mew} using different formulations of the Einstein equations, treatments of initial conditions, and diagnostics, where each study has focused on one smoothing mechanism or the other, as carefully reviewed and critiqued in Ref.~\cite{Ijjas:2023bhh}.  
  These earlier contributions have  informed this paper.    

Our goal here is to perform the first systematic comparative study of inflation and slow contraction, so we have taken a different approach.  We have produced two nearly identical codes that solve the  (3+1)-dimensional Einstein-scalar field equations using the same mean-curvature-normalized, orthonormal tetrad formulation.   This form was chosen because it enables large spatial variations in one or more directions of all freely specifiable physical quantities that characterize the initial spatial hypersurface.    We have also developed a common set of diagnostics to evaluate when initial conditions are outside the perturbative regime of a flat FRW and how smooth and flat the universe must become to satisfy observational constraints.     Hundreds of numerical evolutions, each lasting from a few minutes to a few months, were then run to study how the outcome depends on the different parameters that specify the initial conditions.

Especially important were studies exploring a key difference between inflation and slow contraction.  Namely, smoothing by slow contraction entails a non-linear smoothing mechanism based on ultralocality, a general relativistic effect that appears to be unique to contracting spacetimes.   Ultralocality refers to a condition in which  the spatial derivative terms in the non-linearly coupled partial differential Einstein equations fall exponentially fast during contraction compared to terms containing only time derivatives, converting the partial differential equations to ordinary differential equations.   The ultralocality effect was first conjectured by
Belinski, Khalatnikov and Lifshitz \cite{Belinsky:1970ew} in the 1960s who only considered its application to the case of {\it fast contraction}  ($\varepsilon \le 3$) for historical reasons.  Curiously, in that case, the ordinary differential equations have chaotic solutions that produce  mixmaster behavior that  {\it unsmooths} the universe.  With {\it slow contraction} ($\varepsilon \gg 3$), the ordinary differential equations have an attractor solution that is flat FRW, just what is desired to solve the homogeneity, isotropy and flatness problem.   Inflation, by contrast, occurs during an expanding phase that does not have ultralocal behavior.  This accounts for our finding that inflation is unable to smooth and flatten  the inhomogeneous, anisotropic, and curved initial conditions expected following a big bang.   

The plan of the paper is as follows.  In Sec.~\ref{sec2}, we briefly summarize our numerical relativity schemes and point out the  improvements that have been made compared to  earlier publications. 
In Sec.~\ref{sec3}, we discuss the diagnostics we use in evaluating how far initial conditions  are from flat FRW and how we determine whether smoothing and flattening is achieved.   In Sec.~\ref{sec4}, we
 digress to note several ways in which existing numerical techniques introduce  unavoidable inequities that favor one smoothing mechanism relative to the other.  As it turns out, the inequities all favor inflation and disfavor slow contraction, so they do not qualitatively change our  conclusions.   
 
 Secs.~\ref{sec5} thru~\ref{sec7} present the key results.   In Sec.~\ref{sec5}, we compare two representative examples, one inflation and one slow contraction, both of which have been selected because their initial conditions were, by our diagnostic measures, as far from flat FRW as possible and yet sufficiently limited that smoothing and flattening still occurs.  We have chosen  a negative potential for the scalar field $\phi$ that is exponentially decreasing with 
 $\phi$ for the slow contraction case; and we have chosen  a positive quadratic potential for the inflation case.  Both potentials
 have  been investigated previously by several groups \cite{East:2015ggf,Clough:2016ymm,Aurrekoetxea:2019fhr,Clough:2017efm,Corman:2022alv,Garfinkle:2023vzf} , including us, which is useful for making comparisons.  Although inflation with quadratic potentials is ruled out by experimental bounds on the amplitudes of B-modes, we choose it anyway because it has been suggested  that this is the example that is least sensitive to initial deviations from flat FRW
\cite{Aurrekoetxea:2019fhr,Clough:2017efm}.  We have extended our survey to other potentials and will  report the results elsewhere.  They do not change the qualitative conclusions presented here.

 Based on the diagnostic measures, it is clear that slow contraction is a robust and rapid smoother even for initial conditions that are outside the perturbative  flat FRW, but inflation is not.  In Sec.~\ref{sec6}, we provide new evidence (beyond our earlier work) illustrating how ultralocality leads to smoothing for slow contraction but not for fast contraction.  In Sec.~\ref{sec7}, we then illustrate how the non-ultralocal behavior of inflation evidences itself when initial conditions are far from flat FRW.  We close with a summary  and discussion in Sec.~\ref{sec8}.  

\section{Numerical relativity scheme}
\label{sec2}
We numerically solve the (3+1)-dimensional Einstein-scalar field equations in mean-curvature-normalized, orthonormal tetrad form, as has been successfully implemented in earlier studies of contracting spacetimes \cite{Garfinkle:2008ei,Cook:2020oaj,Ijjas:2020dws,Ijjas:2021gkf,Ijjas:2021wml,Ijjas:2021zyf,Kist:2022mew} and more recently for studies of inflation \cite{Garfinkle:2023vzf}.   The numerical scheme is explained in great detail in the appendix of Ref.~\cite{Garfinkle:2023vzf}.   
 Whereas the existing contraction tetrad code allowed spatial variations along one, two or three spatial directions, the earlier published inflationary tetrad code results  only allowed spatial variations along a single spatial direction \cite{Garfinkle:2023vzf}.  Our first step, therefore, was to write a new inflation tetrad code that allows variations along two or three spatial dimensions.   In the process, a number of small  improvements were made to increase  efficiency and accuracy.  Then, to keep the comparison study as even-handed as possible, we made a mirror version  that uses the same subroutines, functions, algorithms, etc. but which is adapted for contraction.  The study results here are from simulations with initial spatial variations along two directions, as described below.  We found no significant differences from high-resolution simulations with initial variations along one direction.  Simulations of the same resolution
with initial variations along three directions would require at least a thousand times the already significant computer resources used in this study.  However, we have run lower resolution simulations with spatial variations along three directions and confirmed that the qualitative differences between contraction and expansion described in Sec.~\ref{sec7} appear to be maintained.

There are three obvious differences between the inflation and slow contraction software packages that involve only a few lines of code and do not affect efficiency or accuracy:  the time slicing, the scalar field potential, and the initial conditions. 

The slow contraction code uses constant mean curvature time-slicing, 
\begin{equation}
\label{cmctime}
\frac{{\rm d}\ln K}{{\rm d}t} =  -1,
\end{equation}
where the coordinate $t$ runs from 0 to $-\infty$ and  $K\equiv K_a^a$ is the mean curvature, which is equal to $1/3 |H^{-1}|$ in the homogeneous limit.  Note that the mean curvature changes by an exponential factor during slow contraction. Note that $t$ is {\it not} the FRW time coordinate (indicated by $\tau$ in the previous section).  In the homogeneous limit, $|t|$ represents the number of $e$-folds of contraction of the Hubble radius.  

In  the expanding case, the coordinate time $t$ runs from zero to $+\infty$, and the mean curvature only changes by a few orders of magnitude. As a result, a modified time slicing has been found to be necessary \cite{Garfinkle:2023vzf},    
\begin{equation}
\label{newtime2}
\frac{{\rm d}\ln |K|}{{\rm d}t} =  -\mu(t),
\end{equation}
where 
\begin{equation}
\label{def-C02}
\mu(t) \equiv \frac{1}{\tilde{{\cal N}}_{\rm max}(t)},
\end{equation}
$K =-3H$ in the homogeneous limit, and $\tilde{{\cal N}}_{\rm max}(t)$ is the maximum value of the  coordinate lapse $N$ at time $t$ rescaled by the mean curvature divided by $\mu$, $\tilde{\cal N} \equiv {\textstyle \frac13}KN/\mu$ to make it dimensionless. 
As shown in the appendix of Ref.~\cite{Garfinkle:2023vzf}, this time coordinate $t$ then measures the maximum number of $e$-folds of inflation taking place during the simulation.  

As for the scalar field potentials,   a negative potential is required for slow contraction, as noted in the Introduction.  We use a well-studied exponential form,  $V(\phi) = - V_0 \, \exp(\phi/M)$, where $M< M_P$ and $M_P= 2.4 \times 10^{18}$~GeV is the reduced Planck mass. (This study only explores the smoothing phases and not the reheating and latter stages, we do not need to add a minimum to the potential or a bounce  as would be required in the full evolution since we do not get close to that stage.)  In the homogeneous limit, a canonical scalar field rolling down a negative exponential of this form has an attractor solution in which the equation of state $\varepsilon \rightarrow 1/2 M^2$.   For this survey, we limit consideration to values of $M$ between 0.1 and 0.2 (corresponding to attractor solutions with $\varepsilon$ between 13 and 50), which has been shown to be sufficient to smooth the universe \cite{Ijjas:2020dws,Ijjas:2021gkf} (although much larger values are possible and smooth even more robustly and rapidly).  In the case of inflation, a positive potential is required.  We use a well-studied example, $V(\phi) = \frac{1}{2}m^2 \phi^2$, where $m = 3.8  \times 10^{-3} M_P$.  

As a practical matter, we confined the high-resolution studies to spatial variations along one and two directions, which is sufficient to check for consistency and for any non-linear effects that may arise when there are variations along more than one direction.  Extending the survey to high-resolution simulations with initial spatial variations along three dimensions requires parallelism over multiple CPUs and much more expensive simulations.   As a check, though, we have run lower-resolution examples with initial spatial variations along three dimensions and found no qualitative differences in the outcomes and from the results reported in Sec.~\ref{sec7}.

The initial conditions were set using York's conformal method \cite{York:1971hw} for both inflation and slow contraction to ensure they satisfy the Hamiltonian and momentum constraints.  
 In our implementation, the spatial metric on the initial time slice ($t=t_0$) is  set to be  conformally-flat,  
$\gamma_{ij}(t_0,\vec{x}) = \psi^4(t_0,\vec{x})\delta_{ij}$,
where $\psi$ denotes the conformal factor. 
The components of the spatial 3-curvature tensor $\bar{n}_{ab}, \bar{A}_b$ and the tetrad vector components ${\bar{E}{}_a}^i$ then satisfy: 
\begin{alignat}{2}
\label{eq3}
&\bar{n}_{ab}(t_0,\vec{x}) &&= 0 ,  \\
&\bar{A}_b(t_0,\vec{x}) &&= -2\psi^{-1}{\bar{E}{}_b}^i\partial_i\psi,\label{eq4} \\
& {\bar{E}{}_a}^i(t_0,\vec{x}) &&= \psi^{-2}(K_0/3)^{-1}{\delta_a}^i.
\end{alignat}
where 
$\bar{A}_b \equiv {\textstyle \frac12} \varepsilon_{b}{}^{cd}\bar{N}_{cd}$
is the antisymmetric part of $\bar{N}_{ab}$ (the nine intrinsic spatial curvature variables) and $\bar{n}_{ab} \equiv \bar{N}_{ab} - \varepsilon_{ab}{}^c \bar{A}_c$ is the symmetric part.  A  ``bar' on top of any variable corresponds to rescaling by the mean curvature ({\it i.e.}, dividing by $K/3$).  Early alphabet indices ($a$, $b$, {\it etc.}) are frame indices and middle indices   ($i$, $j$, {\it etc.})  are coordinate indices.

There remains the freedom to specify the initial scalar field distribution $\phi(t_0,\vec{x})$, the conformally  re-scaled initial scalar field velocity 
$\bar{Q}$,
and the divergence-free (transverse, trace-less) part of the conformally-rescaled shear tensor, $\bar{Z}_{a b}^{TT} (t_0,\vec{x})\equiv \psi^6\bar{\Sigma}^0_{ab}$, parameterized as follows: 
\begin{equation}
\tensor*{\bar{Z}}{_a_b^{TT}} = \left({ \textstyle \frac{K_0}{3}}\right)^{-1}
{ 
\renewcommand*{\arraystretch}{1.3}
\begin{pmatrix}
b_2 +  c_2 \cos{\left({\textstyle \frac{y}{L}}+\alpha_y\right)} &
&  \kappa &
&  c_1 \cos{\left({\textstyle \frac{y}{L}}+\alpha_y\right)} + a_3  \\
\kappa &{\;}
&   b_1+  a_1 \cos{\left({\textstyle \frac{x}{L}}+\alpha_x\right)} &
&  a_2 \cos{\left({\textstyle \frac{x}{L}}+\alpha_x\right)} + c_3  \\
c_1 \cos{\left({\textstyle \frac{y}{L}}+\alpha_y\right)} + a_3 &
&   a_2 \cos{\left({\textstyle \frac{x}{L}}+\alpha_x\right)} + c_3 & {\;}
& - \tensor*{\bar{Z}}{^{TT}_1_1} - \tensor*{\bar{Z}}{^{TT}_2_2}
\end{pmatrix}},
\label{ZZ2}
\end{equation}
where $a_1, a_2,a_3, b_1, b_2, c_1, c_2, c_3, \alpha_x, \alpha_y$ and $\kappa$ are constants; 
\begin{align}
\begin{split}
\label{QQ2}
\bar{Q} &=\left({\textstyle \frac{K_0}{3}}\right)^{-1}\times \Big(f_0 \cos{\left({\textstyle \frac{m_0}{L}} x + d_0\right)} 
+ f_2 \cos{\left({\textstyle \frac{m_2}{L}} y + d_2\right)} + Q_0
 \Big) \\
\phi &= f_1 \cos{\left({\textstyle \frac{m_1}{L}} x + d_1\right)} + 
f_3 \cos{\left({\textstyle \frac{m_3}{L}} y + d_3\right)} +
\phi_0,
\end{split}
\end{align}
where $Q_0, \phi_0, f_0, f_1,f_2, f_3,  m_0, m_1,  m_2, m_3,  d_0, d_1, d_2, d_3$ are constant and denote the mean value, the amplitude, the mode number and the phase of the initial  velocity and field distribution, respectively, and $K_0$ is the initial mean curvature. 

The choice of cosine reflects the fact that, for the numerical simulation, we must choose periodic boundary conditions $0\leq x, y  \leq2\pi L \Theta_0$ with  $2\pi L\Theta_0$ identified as the `box side-length,' where $\Theta_0$ is the initial inverse mean curvature and  $L \ge 1$ is a positive integer.  With periodic boundary conditions, the mode numbers ($m_i$) must be fixed integers, so the box side-length must be co-moving, {\it i.e.} with a box side-length scaling as $a(t)$.  

For slow contraction, the scale factor shrinks very little over the course of the simulation.  We start with an initial inverse mean curvature $\Theta_0$ that is larger than the present Hubble radius $\sim {\cal O}(10^{26})$ but have the freedom to choose the box side-length to be yet larger still.   
This ensures that the box encompasses from the start a spatial volume large enough to evolve through slow contraction, bounce and expansion into the size of the observable universe today. Then we choose  the number of grid points such that the initial grid spacing is as small as possible compared to the initial mean curvature since our goal is to study the effects of introducing large amplitude  spatial variations that range over scales including ones smaller than the initial mean curvature.  The optimal choice is to choose $L=1$ and maximize the number of grid points as much as computer resources allow.  If  simulations end with the entire box being smooth and flat, this indicates that slow contraction has successfully resolved the homogeneity, isotropy and flatness problems beginning from an initial state that is outside the perturbative regime of flat FRW spacetimes.  
  
Inflation simulations are quite different.  The scale factor $a(t)$ expands by an exponential factor over the course of the simulation and, hence, so does the box side-length (when expressed in physical units).  We start with an inverse mean curvature scale $\Theta_0$ that is $1/2 \pi L$ times the box side-length.  Physically, this length is microscopic compared to the present Hubble radius in physical units.   (More precisely, $\Theta_0$ is of order the Hubble radius at the beginning of inflation.)  We choose a grid spacing that is smaller than $\Theta_0$ because we want to investigate the effects of initial  spatial variations that vary over length scales smaller than $\Theta_0$ (the kinds of variations expected following a big bang).   During the early stages of the simulation before accelerated expansion begins, the Hubble radius can grow faster than $a(t)$,  depending on the initial conditions, and, as a result, can grow relative to the box size by a modest factor.  We want to choose $L$ large enough to follow that stage of the evolution.  Then, once inflation starts, the situation reverses: the Hubble radius grows much more slowly than $a(t)$ and, hence, shrinks relative to the box size.     Well before the end of inflation, the Hubble radius becomes smaller than the grid spacing.  The best compromise for following the evolution on scales of order the Hubble radius during both the early and late stages  is to choose $L \gg1$ and maximize the number of grid points as much as computational resources allow
(see related discussion in Sec.~\ref{sec4}).

We refer the reader to Ref.~\cite{Ijjas:2023bhh} for a review of how to properly apply the York method using an elliptic solver in order to obtain initial conditions for the conformal factor $\psi$ consistent with the Hamiltonian and momentum constraints even for conditions outside the perturbative regime of  flat FRW spacetimes and how to evolve the combined hyperbolic-elliptic system of Einstein-scalar partial differential equations.

\section{Diagnostics}
\label{sec3}

 The goal of our survey is to determine for inflation and slow contraction how far the inhomogeneous, anisotropic and curved  initial conditions can be from the target final state -- a flat FRW spacetime dominated by the scalar field ($\Omega_{\phi}=1$) -- and still reach the target by the end of the smoothing phase.   
   
Four objective quantitative tests are:
 \begin{enumerate}
 \item {\bf $|\Omega_{\phi}^0|$ test:}   
 A robust smoothing mechanism should have dynamical properties that enable the associated stress-energy  component ({\it e.g.}, due to the scalar field in our cases)  to overtake the other degrees of freedom for a sustained period long enough to smooth and flatten the universe.   One test for robustness is, therefore, showing that smoothing occurs beginning from $|\Omega_{\phi}^0|<<1$.  We screen out cases where this initial condition is not satisfied.   

\item {\bf $\lambda/(2 \Theta_0)  < 1$}:  A second test is whether smoothing occurs if the initial spatial variations include modes whose wavelengths are comparable to the initial inverse mean curvature $\Theta_0$  (which, in a homogeneous universe would be the Hubble radius).  Otherwise, the initial conditions would be ones for which the universe is relatively smooth on the scale of the Hubble radius at the very start, which is neither expected physically nor the kind of condition we aim to test.  Due to the periodic boundary conditions, the initial spatial variations given in Eqs.~(\ref{ZZ2}) and~(\ref{QQ2}) are expressed as sums of fourier modes with different wavelengths $\lambda_i= 2 \pi L \Theta_0/m_i$ commensurate with the periodic box side-length $2 \pi L \Theta_0$, where $m_i = m_0, \, m_1, \, m_3, \ldots$ are integers.  
%NOTE:  what is called "box" in the code is the integer L.  The actual boxsize is $2 \pi L \Theta_0$.
With this parameterization, this  second test reduces to checking  if  $ \pi L /m_i  <1$ for one or more of the $m_i$.  Whereas the condition on  $|\Omega_{\phi}^0| \ll 1$ is not so difficult to obtain for either inflation or so contraction, this second test is more challenging, especially for inflation where the evolution is not ultralocal.  
 
\item {\bf Weyl curvature $\hat{\cal C}$ and Chern-Pontryagin  invariant $\hat{\cal P}$}:  These two scalars, derived from the Weyl curvature tensor, were first proposed as good measures of inhomogeneity, anisotropy and curvature in Ref.~\cite{Ijjas:2023bhh} because they are independent of the formulation, gauge or frame.  

The  conformal Weyl curvature tensor $\tensor{\cal C}{_\mu_\nu_\rho_\sigma}$ is the  trace-free part of the Riemann curvature tensor.  The scalars derived from it   are
\begin{equation}
\label{def-weylscalar}
{\cal C} \equiv \tensor{\cal C}{^\mu^\nu^\rho^\sigma}\tensor{\cal C}{_\mu_\nu_\rho_\sigma}
\end{equation}
 and
\begin{equation}
\label{def-CPscalar}
{\cal P} \equiv  {}^{\ast}\tensor{\cal C}{^\mu^\nu^\rho^\sigma}\, \tensor{\cal C}{_\mu_\nu_\rho_\sigma},
\end{equation}
where
\begin{equation}
\label{def-dualWeyl}
{}^{\ast}\tensor{\cal C}{_\mu_\nu_\rho_\sigma}
\equiv \frac12 \tensor{\chi}{_\mu_\nu^\tau^\zeta}\,\tensor{\cal C}{_\tau_\zeta_\rho_\sigma},
\end{equation}
with $ \tensor{\chi}{_\mu_\nu_\tau_\zeta}\equiv - \sqrt{|-g|}\tensor{\varepsilon}{_\mu_\nu_\tau_\zeta}$ being the totally anti-symmetric Levi-Civita 4-form and $\tensor{\varepsilon}{_\mu_\nu_\tau_\zeta}$ being the Levi-Civita tensor.
The two curvature invariants can be expressed in terms of two  spatial 3-tensors $E_{ab}$ and $H_{ab}$ corresponding to the electric and magnetic components of the Weyl curvature tensor, as detailed in \cite{Ijjas:2023bhh}:
\begin{eqnarray}
\label{Weyl-invariant}
\label{def2-weyl}
{\cal C} &=& 8\Big( E_{ab}E^{ab} - H_{ab}H^{ab}\Big),\\
\label{def2-cp}
{\cal P} &=& 16 E_{ab}H^{ab}.
\end{eqnarray}
Notably, $E_{ab}$ depends directly on the shear tensor $\Sigma_{ab}$ and $H_{ab}$ depends on contractions of $\Sigma_{ab}$ with components of the spatial curvature $N_{ab}$ and on the spatial derivatives of  $\Sigma_{ab}$.  This significance of this distinction in testing inflation vs. slow contraction will be discussed in the next section.

For flat FRW, $\hat{\cal C} =\hat{\cal P}=0$.    A third objective test of a smoothing mechanism is how large the mean  $|\hat{\cal C}|$ and $|\hat{\cal P}|$ (averaged over the simulation volume) can be initially and yet have the smoothing mechanism reduce them to $|\hat{\cal C}|<10^{-10}$ and $|\hat{\cal P}|< 10^{-10}$
by the end of the simulation.  Here the hat refers to normalizing by the initial Ricci curvature.   The initial mean values of $|\hat{\cal C}|$ and $|\hat{\cal P}|$ must  both be greater than unity to be significantly outside the perturbative regime of flat FRW.

The upper bounds of $10^{-10}$ on the final mean values  $|\hat{\cal C}|$ and $|\hat{\cal P}|$
are necessary to ensure that the  deviations from flat FRW  after smoothing are negligible compared to the 
quantum-generated curvature  perturbations  generated during the last stages of smoothing.  Those quantum-generated perturbations are supposed to seed the observed temperature fluctuations in the cosmic microwave background.  The observed amplitude of those fluctuations (squared) $(\delta T/T)^2 \sim 10^{-10}$, sets the $10^{-10}$ upper bound on $|\hat{\cal C}|$  and $|\hat{\cal P}|$ after smoothing.

\item {\bf $\sigma^2_{\cal C}$ and $\sigma^2_{\cal P}$}:   A fourth objective test is how large the variances of $|\hat{\cal C}|$ and $|\hat{\cal P}|$ can be initially and yet have smoothing reduce them negligible levels.  Checking the variance in addition to the mean is an important independent test because it screens out special instances where the means are large but nearly homogeneous. As with the means, the variances of $|\hat{\cal C}|$ and $|\hat{\cal P}|$ must  both be greater than unity to be significantly far from flat FRW.
\end{enumerate}

\section{Limitations}
\label{sec4}

Despite efforts at conducting a fair comparison of inflation and slow contraction, such as designing mirror codes and common objective diagnostics, numerical general relativity imposes certain unphysical technical limitations that tend to bias the outcomes.  In the interest of full disclosure and to inspire improvements in future similar studies, we describe these here.  

As it turns out, all the known technical biases are in the direction of favoring inflationary smoothing.   Since our finding is that slow contraction solves the homogeneity, isotropy and flatness problems and inflation does not, our qualitative conclusion is not changed  by these limitations.  If anything, the quantitative advantages of slow contraction are being underestimated.

The biases trace back to the fact that   the shear (anisotropy) $\Sigma_{ab}$ grows faster than spatial curvature $n_{ab}$  in a contracting universe, and the reverse is true in an expanding universe.  Shear and spatial curvature are the key components  that can most effectively block  a scalar field from dominating and thereby prevent smoothing.   A  limitation of our current implementation of the York method used to set initial conditions that satisfy the Hamiltonian and momentum constraints is that 
 the conditions imposed on $\Sigma_{ab}$ and $n_{ab}$  are fundamentally different.
 
In particular, our implementation  places no significant restriction on $\Sigma_{ab}$  but requires $n_{ab}$ to be precisely zero on the entire initial spatial hypersurface.   
  Fixing $n_{ab}=0$ on the initial time-slice advantages inflation.  It does not force $n_{ab}$ to be zero for all time.  In fact, the non-linear evolution of the Einstein equations generates a non-zero $n_{ab}$ within a few time steps.  However, the average curvature will remain zero so that there are necessarily patches where the spatial curvature passes through zero across the simulation.  These patches are places where inflation has no spatial curvature to overcome, which advantages inflation, but only because of artificial limitations of current numerical simulation techniques. 

A similar situation occurs with $\hat{\cal C}$ and $\hat{\cal P}$ and with the electric and magnetic components of the Weyl tensor,  $E_{ab}$ and $H_{ab}$.  The electric component depends on the shear, and so is periodic but otherwise unconstrained.  However, the magnetic component is comprised mostly of terms proportional to  contractions of the shear with $n_{ab}$.  These terms are zero on the initial time-slice and, even as $n_{ab}$ varies in later steps, the magnetic term will tend to pass through zero in patches where $n_{ab}$ passes through zero.  Since ${\cal P} = 16 E_{ab}H^{ab}$, the Chern-Pontryagin invariant $\hat{\cal P}$ does this same. Again, inflation is favored and   slow contraction receives no analogous advantages.

In principle, the York method can be extended to handle metrics with non-zero spatial curvature. This would be a useful technical advance, albeit one that is challenging.   If that were to be achieved, though, the expectation would be that inflation would lose its artificial advantage and smoothing would be more difficult. 
 
Another difference between simulations of inflation versus slow contraction has to do with the box side-length $2 \pi L \Theta_0$.   
As described in Sec.~\ref{sec2}, contraction can be optimally simulated with $L=1$; inflation requires $L$ to be made as large as computational resources allow.  In the inflation case, it is necessary to increase  the integer mode numbers  $m_i$ in the arguments of the cosines in Eq.~(\ref{QQ2}) by a  factor of order $L$.   Otherwise, by increasing the box side-length but keeping the $m_i$ fixed, the initial wavelengths of any spatial variation become much larger than the inverse mean curvature from the very start, an initial condition that is nearly homogeneous over a Hubble volume.  This consideration, which artificially favors inflation,  was not included in cases presented our earlier paper  \cite{Garfinkle:2023vzf}.  This motivated developing the second diagnostics  described in Sec.~\ref{sec3}.

\section{Survey results} 
\label{sec5}

The results of the systematic survey of inflation and slow contraction are unambiguous.  They do not require looking through hundreds of examples.  Rather, it suffices to consider  representative cases.  For inflation, 
we varied the parameters that determine the initial spatial variations of the shear, scalar field gradient, scale field velocity --  both individually and in combinations --  to find examples 
that came closest to satisfying the diagnostic criteria described in Sec.~\ref{sec3} and still smooth and flatten; no examples satisfied all of them.   For slow contraction, we could satisfy them all, and we varied the same parameters to find the initial conditions that most exceed the criteria and yet still smooth. 
Two representative cases are summarized in Table~\ref{Fig1}.   The differences are great both qualitatively and quantitatively.

\begin{table*}[tbp]
\centering
\renewcommand{\arraystretch}{1.75}
\setlength{\tabcolsep}{12pt}
\begin{tabular}{ |c |c|c|  }
 \hline
 & Inflation & Slow Contraction 
 \\ \hline
$ \lambda/|2\Theta|< 1$ &   \cellcolor{red} $\lambda/|2\Theta| = 4.5$ &   \cellcolor{green} $\lambda/|2\Theta| = 0.48$
 \\ \hline
 \multirow{2}{*}{max $\bar{\cal C}$,  max $\bar{\cal P} > 1$} &  \cellcolor{red} $\bar{\cal C}=3.5$ &  \cellcolor{green} $\bar{\cal C}=7\times10^7$\\
 &  \cellcolor{red} $\bar{\cal P}=0.3$ &  \cellcolor{green} $\bar{\cal P}=9\times10^6$
  \\ \hline
   \multirow{2}{*}{$\sigma_C^2 >1$,  $\sigma_P^2 >1$} &   \cellcolor{red} $\sigma_C^2=0.4$ &  \cellcolor{green} $\sigma_C^2=1.1\times10^{13}$\\
 &   \cellcolor{red} $\sigma_P^2=0.003$ &  \cellcolor{green} $\sigma_P^2=3.3\times10^{11}$
  \\ \hline
\end{tabular}
\caption{Results of the three critical tests described in Sec.~\ref{sec3}  to assess whether a smoothing mechanism can solve the homogeneity, isotropy, isotropy problems beginning from initial conditions that are non-perturbatively far from flat FRW.  To qualify, a smoothing mechanism must be able to smooth in cases where the initial conditions satisfy all three criteria.  In the survey, no inflationary example was found that satisfied any of the three conditions and still smoothed and flattened (as indicated by the red boxes); the initial conditions for an example that came closest and still smoothed are shown in the box.  By contrast, a wide range of slow contraction examples satisfied all three conditions by a substantial margin and still smoothed and flattened; the initial conditions for an extreme case in the survey are shown in the green boxes. 
\label{Fig1}}
\end{table*}

\begin{figure}[t]%
    \centering
\includegraphics[width=0.6\linewidth]{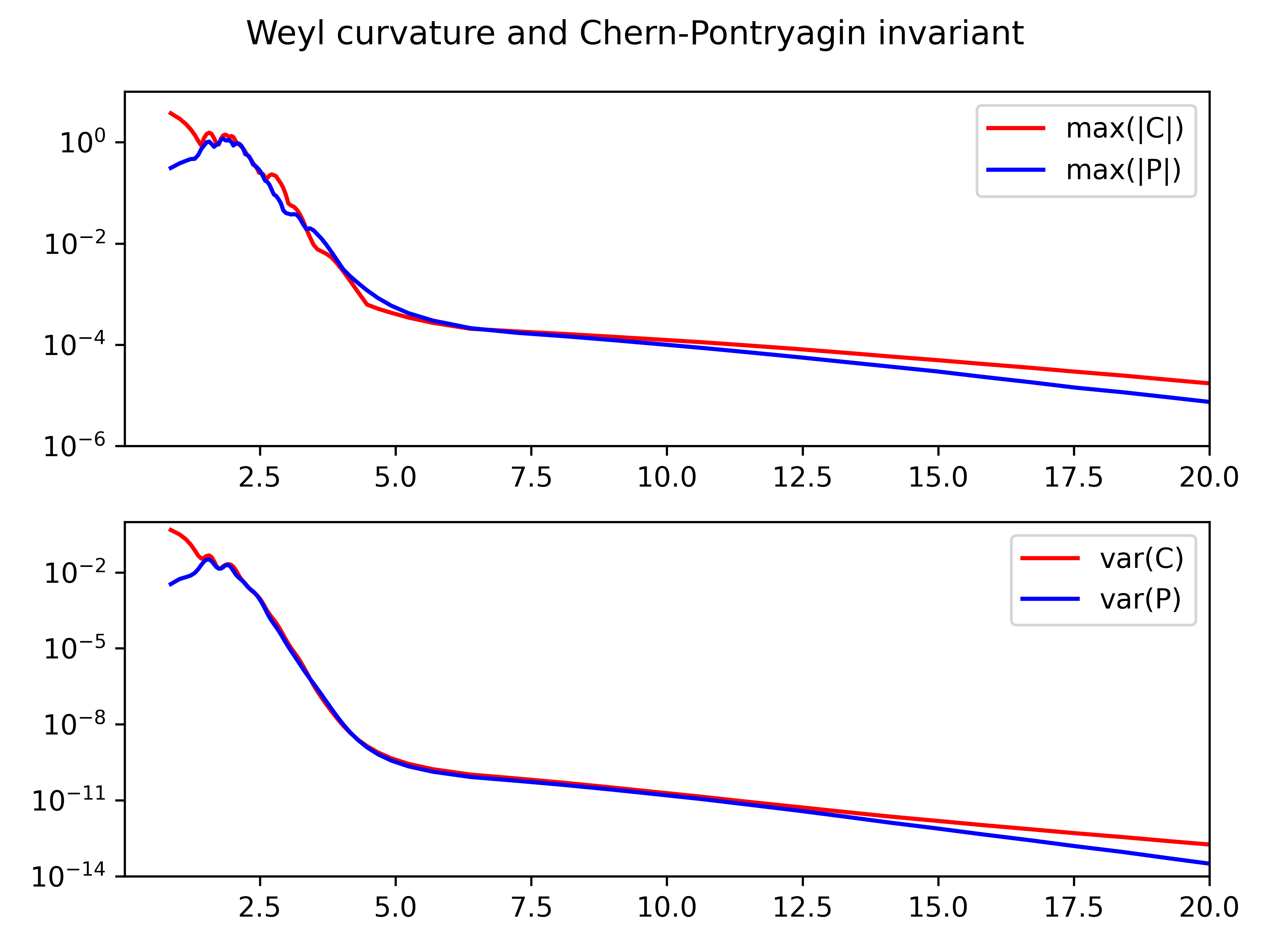}%
    \caption{For  the inflation model described in Table~\ref{Fig1}, the evolution of the max values of $|\hat{\cal C}|$ and $|\hat{\cal P}|$ in the simulation and their variances as a function of the $t$ coordinate (equal to the number of $e$-folds of inflationary expansion).} 
    \label{Fig2}%
\end{figure}

For both inflation and slow contraction, many examples satisfied the first criterion in Sec.~\ref{sec3}, $|\Omega_{\phi}| \ll 1$, including the two representative cases presented in Table~\ref{Fig1}.   However, no inflation example was found that passed the other three tests and still smoothed and flattened, as indicated by the red boxes in Table~\ref{Fig1}.  The initial conditions for the case that came closest are detailed in the figure.    
Its parameter values, as defined in Eqs.~(\ref{ZZ2}) and~(\ref{QQ2}), are:
%jtest13 -- had some spiking at beginning but overcomes it; complex orbit plot that converges but barely
% in fact C and P do not reach the 10^(-10) level.  jtest18 smoothed but did not spike as much.
\begin{align}
\label{init-inflation}
& K_0 = -3;\; a_1=1.7,\;a_2=3.0,\; a_3= 1.5,\; b_1=1.8,\; b_2=-1.5,
\\
& c_1=5.0,\; c_2=5.0,\; c_3=10.0,\; \kappa=0.01; \; \alpha_x=\alpha_y=0.0 \\
& f_1 = 0.24,\; m_1=49,\; d_1 =-1.05,\; f_3 = 0.29,\; m_3=42,\; d_3 =0.0,\; \phi_0=25;\\
& f_0 =0.44,\; m_0 =6,\; d_0=-1.57,\; f_2 = 0.18,\; m_2 =5,\; d_2=0.0,\; Q_0 = -5.0.
\end{align}   
The box side-length is $2 \pi L \Theta_0$ where $\Theta_0 = 3/K_0$ and  $L=70$, which is sufficient to ensure that the box is larger than the Hubble volume throughout the simulation, as required.  In the quadratic potential of the scalar field, $m^2 = 1.43 \times 10^{-5}$ in reduced Planck units.
Fig.~\ref{Fig2} shows the evolution of the average values of $\hat{\cal C}$ and $\hat{\cal P}$ in the simulation as a function of $t$ (whose value equals the number of $e$-folds of expansion) is shown for the first 20 $e$-folds.  It is apparent that the initial conditions fail two critical tests in that the initial values of $\hat{\cal P}$and its  variance are significantly less than one.   (Note also that the max values do not reach the smoothness bound of $10^{-10}$ within the first 20 $e$-folds.)

%Figure2
%%

%Figure3
\begin{figure}[b]%
    \centering
\includegraphics[width=0.6\linewidth]{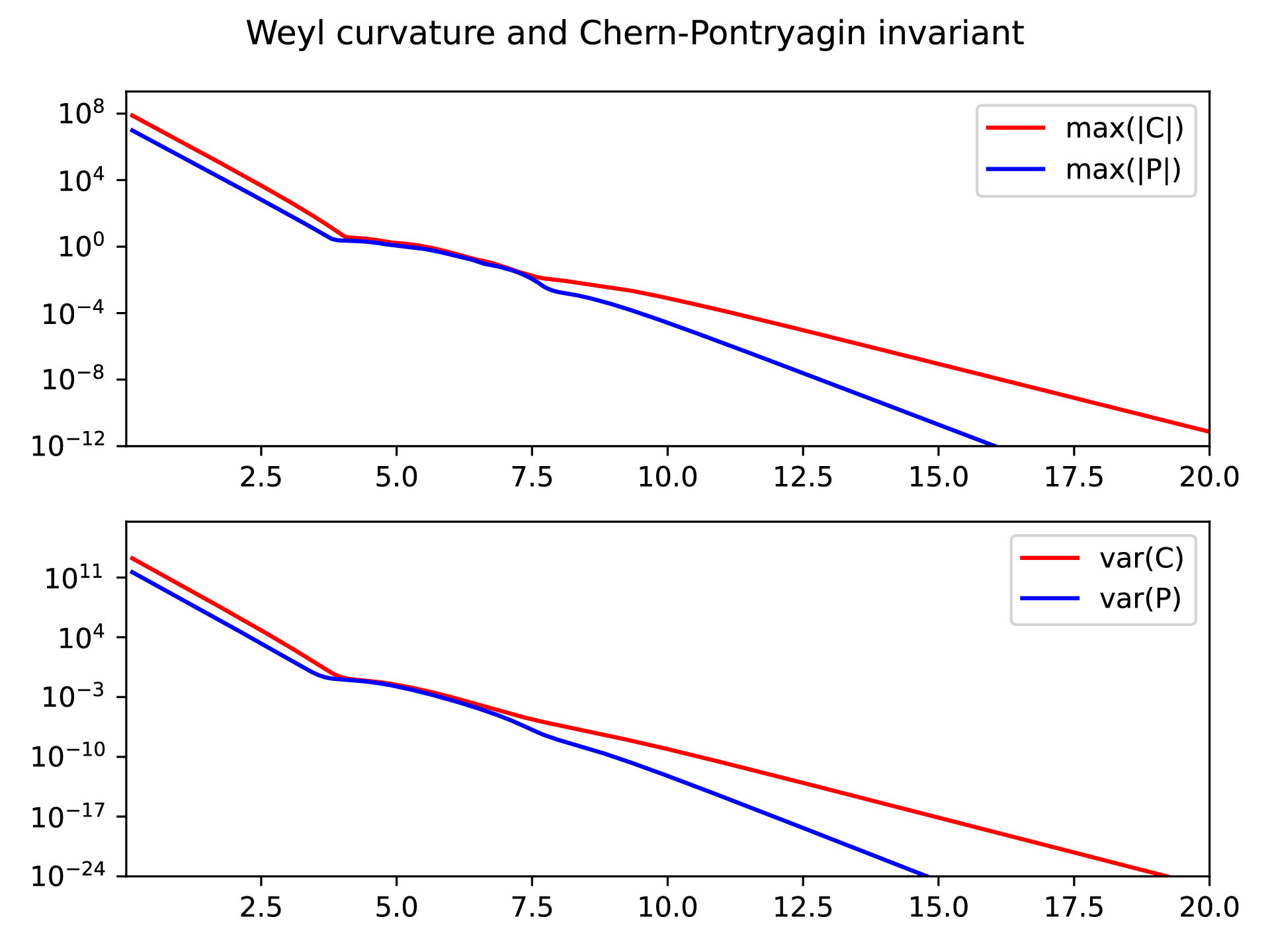}%
    \caption{For the  contraction model described Fig.~1, the evolution of the max values of $|\hat{\cal C}|$ and $|\hat{\cal P}|$  in the simulation and their variances as a function of the $t$ coordinate (equal to the number of $e$-folds of contraction of the Hubble radius).} 
    \label{Fig3}%
\end{figure}

By contrast, many slow contraction models satisfied all four tests in Sec.~\ref{sec3} by a substantial margin.   Table~\ref{Fig1} gives the details for a slow contraction example in the survey that smoothed despite initial conditions were among those  furthest from flat FRW and yet smoothed.  Its parameters are:
\begin{align}
\label{init-contraction}
& K_0 = -3.0;\; a_1=1.02,a_2=1.8,\; a_3= 0.9,\; b_1=1.08,\; b_2=-0.9,
\\
& c_1=3.0,\; c_2=3.0,\; c_3=6.0,\; \kappa=0.01; \; \alpha_x=\alpha_y=0.0 \\
& f_1 = 0.6,\; m_1=4,\; d_1 =-1.05,\; f_3 = 0.66,\; m_3=3,\; d_3 =0.5,\; \phi_0=0.0;\\
& f_0 = 0.6,\; m_0 =5,\; d_0=-1.57,\; f_2 = 0.72,\; m_2 =2,\; d_2=0.0,\; Q_0 = 2.0.
\end{align}  
The potential coefficient is $V_0=0.1$ in units of $H^2 M_P^2$. 
The box side-length is $2 \pi L \Theta_0$ with  $L=1$, which is sufficient to ensure that the box is larger than the Hubble radius throughout contraction, as required.  In the exponential potential, $M =0.1$ in reduced Planck units, which has an attractor solution with equation-of-state $\varepsilon=50$.  
Fig.~\ref{Fig3} is the  complement to Fig.~\ref{Fig2} for the case of contraction: the evolution of the max values of $|\hat{\cal C}|$ and $|\hat{\cal P}|$ in the simulation and their variances as a function of $t$ (which corresponds to the number of $e$-folds of contraction of the Hubble radius).  The criteria that the initial values of these four quantities are greater than one is exceeded by an exponential factor and they 
 reach  the smoothness bound within  the first 20 $e$-folds despite the extremely inhomogeneous initial conditions.

The central conclusion of the survey, as illustrated by these two examples, is  that slow contraction solves the homogeneity, isotropy, and flatness problems and that inflation is problematic.  This
 is consistent qualitatively with earlier individual cases studies by us \cite{Garfinkle:2008ei,Cook:2020oaj,Ijjas:2020dws,Ijjas:2021gkf,Ijjas:2021wml,Ijjas:2021zyf,Kist:2022mew,Garfinkle:2023vzf,Ijjas:2023bhh}.  However, now one can point to the results of hundreds of runs that systematically explore a wide range of initial conditions, and one can    
measure quantitatively how powerful of a smoother slow contraction really is, far exceeding the requisite limits  on  the average Weyl curvature $\hat{\cal C}$ and the Chern-Pontryagin invariants $\hat{\cal P}$ and their variances by many orders of magnitude!   

We have suggested that ultralocality 
is the reason for the extraordinary smoothing power of slow contraction in Refs.~\cite{Ijjas:2021gkf,Ijjas:2023bhh}, and its absence in an expanding universe is what makes inflation problematic.  In the following two sections, we will provide examples of supporting evidence stemming from the survey.

\section{Ultralocality: slow contraction vs. fast contraction}
\label{sec6}

%Figure4
\begin{figure}[b]%
    \centering
\includegraphics[width=0.6\linewidth]{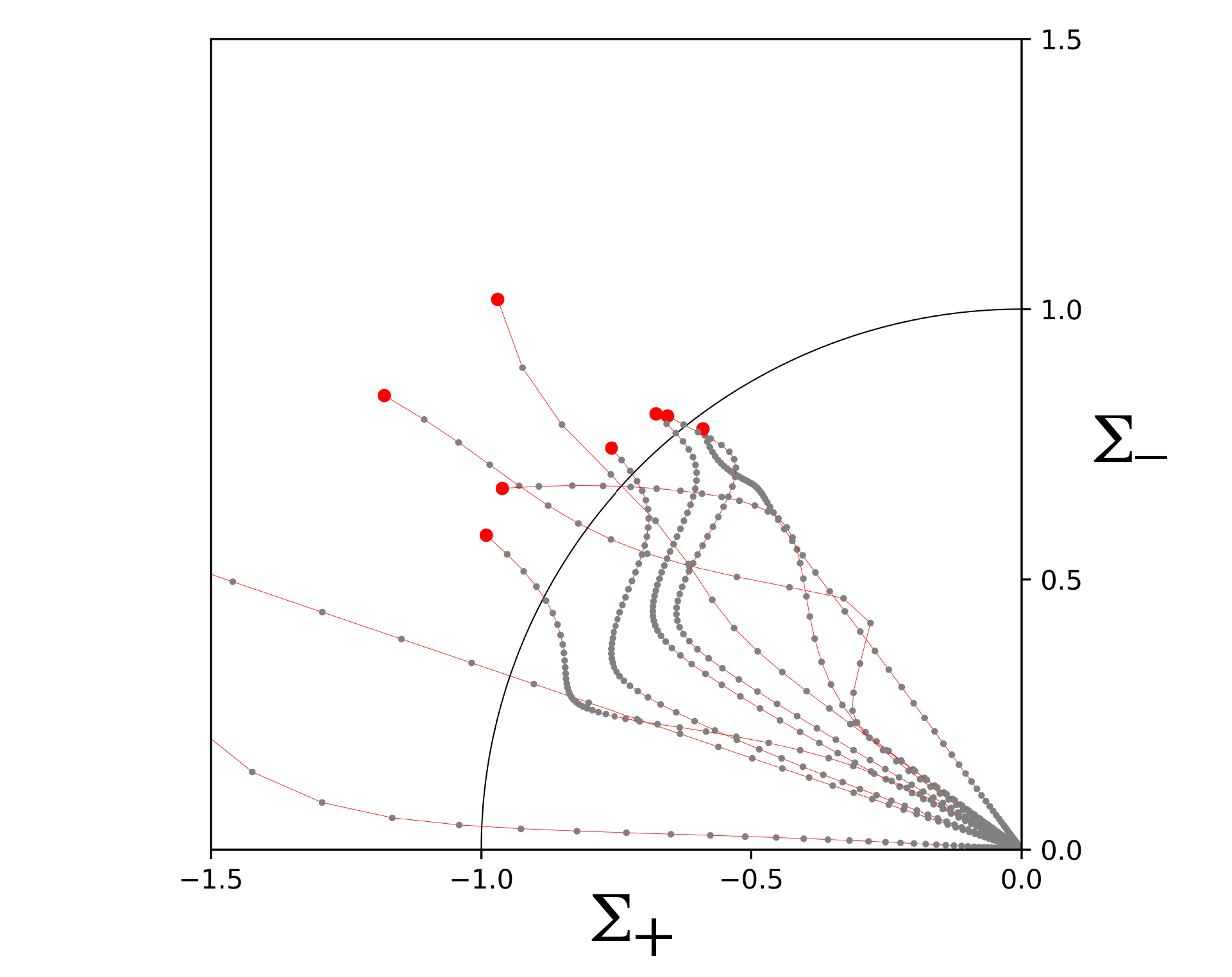}%
    \caption{The state space orbit for the contraction model described in Sec.~\ref{sec4} and Table~\ref{Fig1}  in the $(\bar{\Sigma}_+, \bar{\Sigma}_-)$ showing worldines for ten different spatial points in the simulation.  The small solid red circles correspond to the initial condition at that point. The thin red curves show the trajectory.  The grey dots show indicate the conditions at successive time steps along the trajectory. The quarter circle corresponds to the vacuum Kasner solution, and the center of the circle corresponds to the target flat FRW.}
    \label{Fig4}%
\end{figure}

The notion of ultralocality was introduced  by  Belinski, Khalatnikov and Lifshitz in Ref.~\cite{Belinsky:1970ew} (BKL) who conjectured that, in contracting vacuum space-times, spatial derivatives in the equations of motion become small compared to the time derivatives.   Removing spatial gradients reduces the Einstein-scalar field equations from a system of partial differential equations to a system of ordinary differential equations.  In more recent work, it has been demonstrated through numerical relativity simulations that  this ``ultralocal'' behavior also occurs when there are matter sources or a free scalar field \cite{Berger:1998vxa,Lim:2009dg,Garfinkle:2020lhb}. In all but the free scalar case,  shear grows to dominate the evolution which then triggers chaotic mixmaster behavior.  The chaotic nature results in an uncontrollable growth of inhomogeneity.  The free scalar case is borderline in this sense; in the homogenous limit, the ratio of anisotropy to scalar field energy density approaches a constant.

Ultralocality has also been demonstrated in numerical relativity simulations of  slow contraction  \cite{Ijjas:2021gkf,Ijjas:2023bhh} where the scalar field is self-interacting through a negative potential energy.  If the potential is sufficiently steep (corresponding to $\varepsilon \gg 3$),  the system of ordinary differential equations has an attractor solution that drives widely separated points in space to a common fixed point: flat FRW.  Because different spatial points begin with different initial conditions, they start with different deviations  from flat FRW and evolve at different rates towards it.   However, because the atttractor is a strong one, even points that were not causally connected at the beginning of slow contraction end up reaching the flat FRW condition shortly after it begins and rather rapidly, within twenty or so $e$-folds of slow contraction of the Hubble radius for a wide range of conditions.  

As has been noted in Refs.~\cite{Ijjas:2021gkf,Ijjas:2023bhh}, this novel smoothing behavior contradicts the standard lore that the causal connectivity and causal interaction is essential for homogenizing and isotropizing.    The best that is possible is a local causal smoother, according to this lore.   Yet, with ultralocal smoothing, the first step is to disconnect neighboring points in spacetime  by shrinking the spatial derivative terms in the equations of motion until they are negligible  and relying on attractor behavior to drive the points independently to the same desired common flat FRW endpoint.  In this sense, the smoothing by slow contraction is universal and acausal.

%Figure5
\begin{figure}[b]%
    \centering
\includegraphics[width=0.8\linewidth]{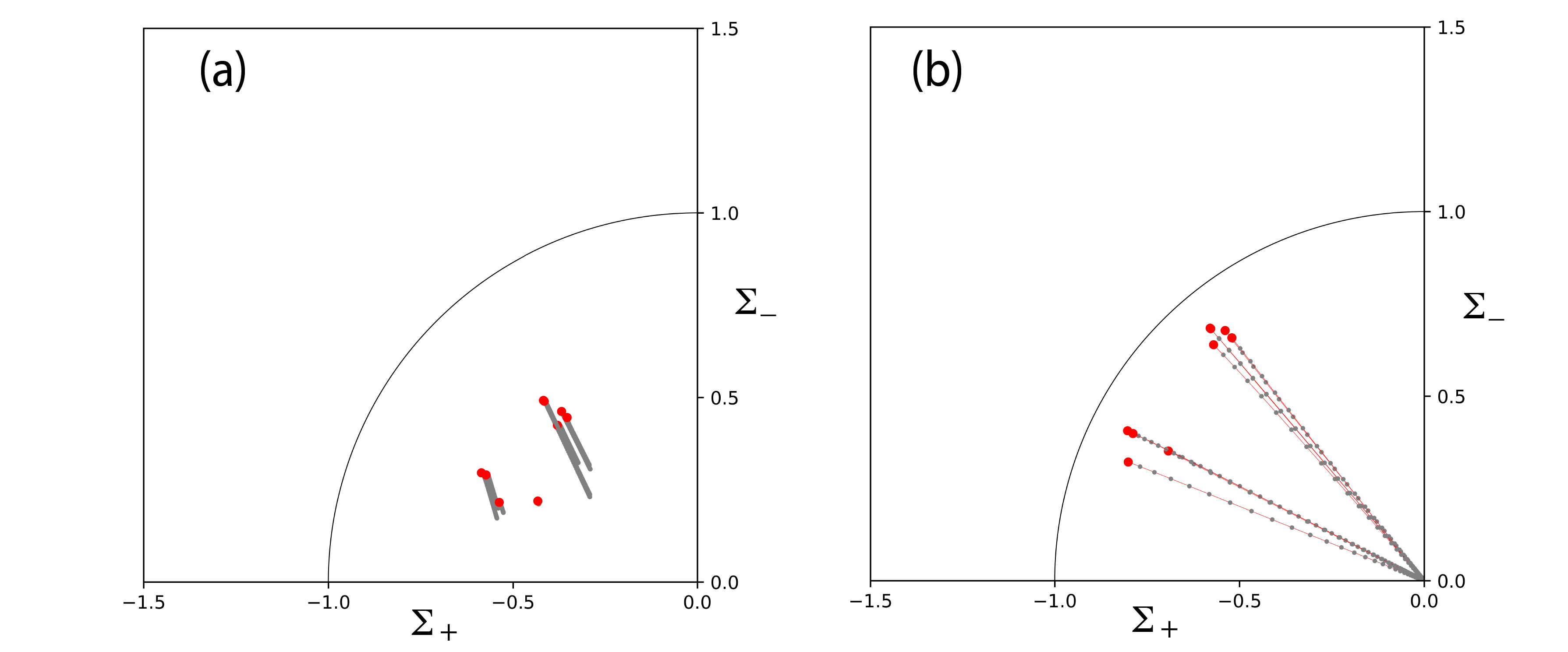}%
    \caption{The state space orbit for two simulations beginning with the same initial conditions.  The only difference is that the scalar field is free in (a), which is a case of fast contraction,  and the scalar field has a negative potential in (b), corresponding to a case of  slow contraction.}   
    \label{Fig5}%
\end{figure}

Fig.~\ref{Fig4}  illustrates this behavior for the slow contraction model described in the previous section by showing the time-evolution of the shear tensor $\bar{\Sigma}_{ab}$ using a standard  state space orbit plot.  The axes are  $(\bar{\Sigma}_+, \bar{\Sigma}_-)$ where $\bar{\Sigma}_{+} = \textstyle{\frac12}(\bar{\Sigma}_{11} + \bar{\Sigma}_{22})$ and $\bar{\Sigma}_{-} =\frac{1}{2\sqrt{3}}(\bar{\Sigma}_{11} - \bar{\Sigma}_{22})$.  
The $\bar{\Sigma}_{\pm}$ are normalized such that the unit circle ($\bar{\Sigma}_+^2+\bar{\Sigma}_-^2 =1$) corresponds to the vacuum Kasner solution.  The center of the circle corresponds to the flat FRW solution.  For simplicity, we show only  one quadrant.   Each trajectory corresponds to a particular spatial point in the simulation, where the points are evenly spread across the box; the red circles represent starting points.  
Trajectories that begin at or  beyond the quarter circles  typically have  $\hat{C}\gg 1$ and $\hat{P}\gg1$.  In this case, many trajectories start significantly outside the circle, consistent with the large mean values reported in Table~\ref{Fig1}.  But each trajectory, traveling over different paths and taking different amounts of time,   is drawn rather directly to the flat FRW point at the center of the circle, even for these extreme initial conditions.  This behavior is a sign of robust and rapid smoothing and is characteristic of the combination of ultralocality and slow contraction.  

%Figure6
\begin{figure}[b]%
    \centering
\includegraphics[width=0.8\linewidth]{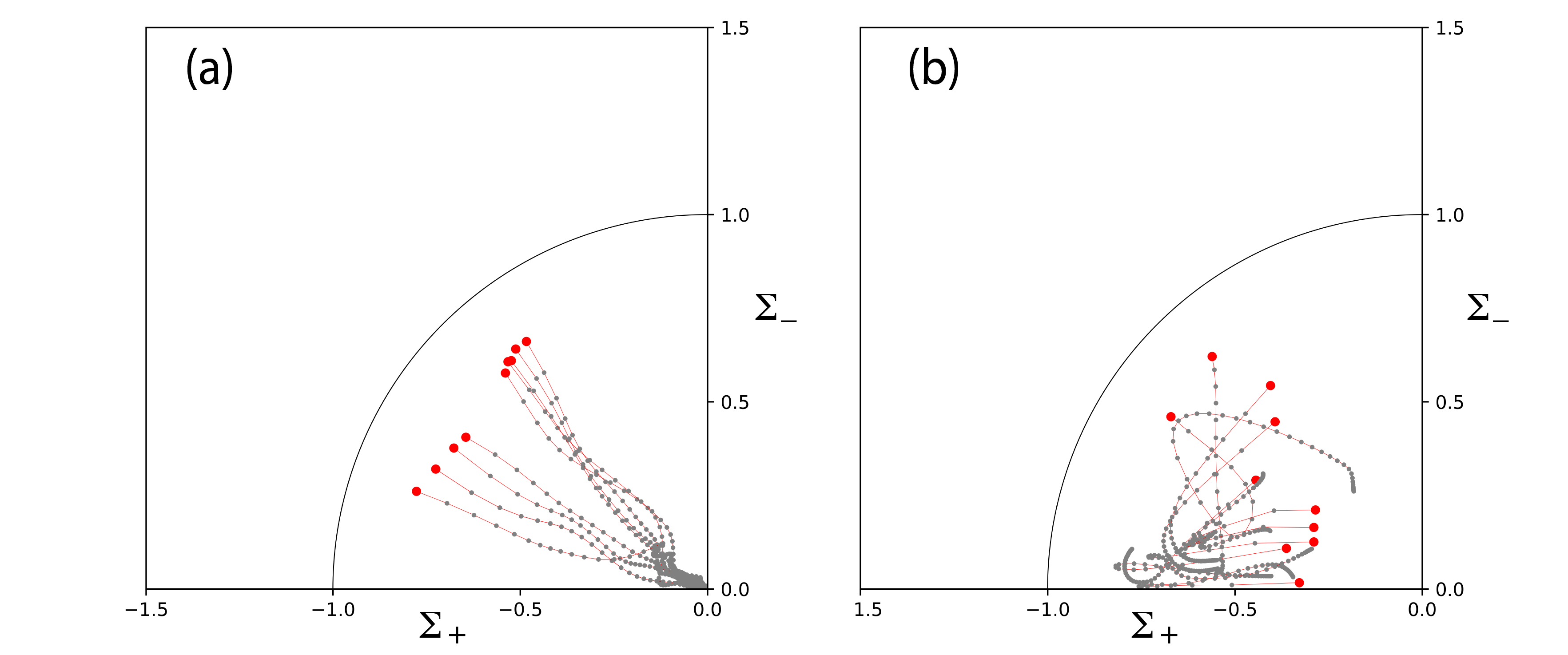}%
    \caption{The state space orbits for two simulations of inflation.   The orbits in (a) are for the inflationary case described in Sec.~\ref{sec5} that comes closest to satisfying the initial conditions criteria delineated in Sec.~\ref{sec3} and still smooths.   The trajectories are inward and reach the center of the circle corresponding to flat FRW.  The plot in (b) is an inflationary case with initial conditions somewhat closer to satisfying the criteria but which fails to smooth.  The trajectories do not reach the center and some even turn away from flat FRW.   
    }   
    %jtest13 and jtest34
    \label{Fig6}%
\end{figure}

One might wonder if ultralocality is all that is required.  
Fig.~\ref{Fig5} shows that the answer is no.  The figure illustrates the difference in the state space orbit plots for two examples of contraction.  Because both are contracting, both evolutions become ultralocal.  In the first case, the scalar field is free ($V(\phi)=0$), corresponding to $\varepsilon=3$; in the second case,  
 the scalar field  has a negative potential with an attractor solution with  $\varepsilon=13$.  Even though the evolution becomes ultralocal in both cases and the starting points are inside the circle (so relatively mild initial conditions compared to Fig.~\ref{Fig4}), the outcomes are quite different.  The trajectories never reach flat FRW (the center of the circle) in the case of fast contraction, but rapidly converge to  flat FRW for the case of slow contraction.  In other words, ultralocality alone does not solve the homogeneity, isotropy and flatness problems, but ultralocality combined with $\varepsilon \gg 3$ clearly can.  (N.B., as detailed in the phase diagrams shown in Ref.~\cite{Ijjas:2021gkf}, the crossover in $\varepsilon$ from non-smoothing to smoothing occurs somewhere between $\varepsilon = 3$ and 13, depending on the initial conditions.) 
 
  As emphasized in Ref.~\cite{Ijjas:2023dnb}, the combination of ultralocality and $\varepsilon \gg 3$ drives the Weyl curvature to zero at spacetime points that are causally disconnected.  Penrose~\cite{Penrose:1979azm} earlier pointed out that this condition is essential for explaining why the early universe is FRW, but he had no physical explanation for it.   Penrose went so far as to propose  a new law of physics (the Weyl Curvature Hypothesis) that initial singularities must have zero Weyl curvature.  However,  the Hypothesis is not only {\it ad hoc}, but it is unlikely to be valid  following a big bang once quantum effects are included.   If the universe passes through a phase with strong quantum gravitational effects following a bang, quantum gravity fluctuations create large random fluctuations in spacetime.   The same criticism applies to slow contraction if continued to such high densities that  quantum gravity effects become large. To fully replace the Weyl Curvature Hypothesis with a physical mechanism, a third essential element is a smooth transition from contraction to expansion (a.k.a. ``non-singular bounce'') at low densities where quantum gravity effects are negligible.

\section{Ultralocality: slow contraction vs. inflation}
\label{sec7}

We have shown above that ultralocality -- the rapid shrinking of spatial derivative terms in the Einstein-scalar field equations in a contracting universe -- plays a central role in solving the homogeneity, isotropy and flatness problems with slow contraction.  This also suggests the reason why inflation is problematic --   the spatial gradient terms in the Einstein-scalar field equations grow.   Heuristic arguments explaining how inflation smooths assume the inflaton field has already dominated the evolution.  But if the initial conditions are outside the perturbative regime of flat FRW, as expected, for example, following a big bang, there is  spatially varying shear and curvature with significant spatial gradients that can dominate the evolution and block inflation if those gradients grow.  

The state space orbits in Fig.~\ref{Fig6} exemplify the problem.  On the left are the orbits for the inflationary example described in Table~\ref{Fig1} that comes closest to satisfying the four criteria in Sec.~\ref{sec3} and yet smooths.  The smoothing is indicated by the convergence of trajectories at the center of the Kasner circle.  Note that the starting points are all inside the Kasner circle and much closer to the center compared to the slow contraction case.  Also the trajectories are converging in way that becomes increasingly twisty as they approach the center, which differs from the slow contraction examples.

The right is an inflationary example with initial conditions that come somewhat closer to satisfying the criteria.  In this case, though, smoothing fails.   The opposite of ultralocal behavior occurs. Spatial gradients are growing sufficiently rapidly that inflation cannot take hold.  Some trajectories end up turning  away from the center.  And this occurs when the starting points (indicated by the solid red circles) are well inside the Kasner circle indicating that the initial conditions are not extreme, as they are in the slow contraction example above.

\begin{figure}[t]%
    \centering
\includegraphics[width=0.9\linewidth]{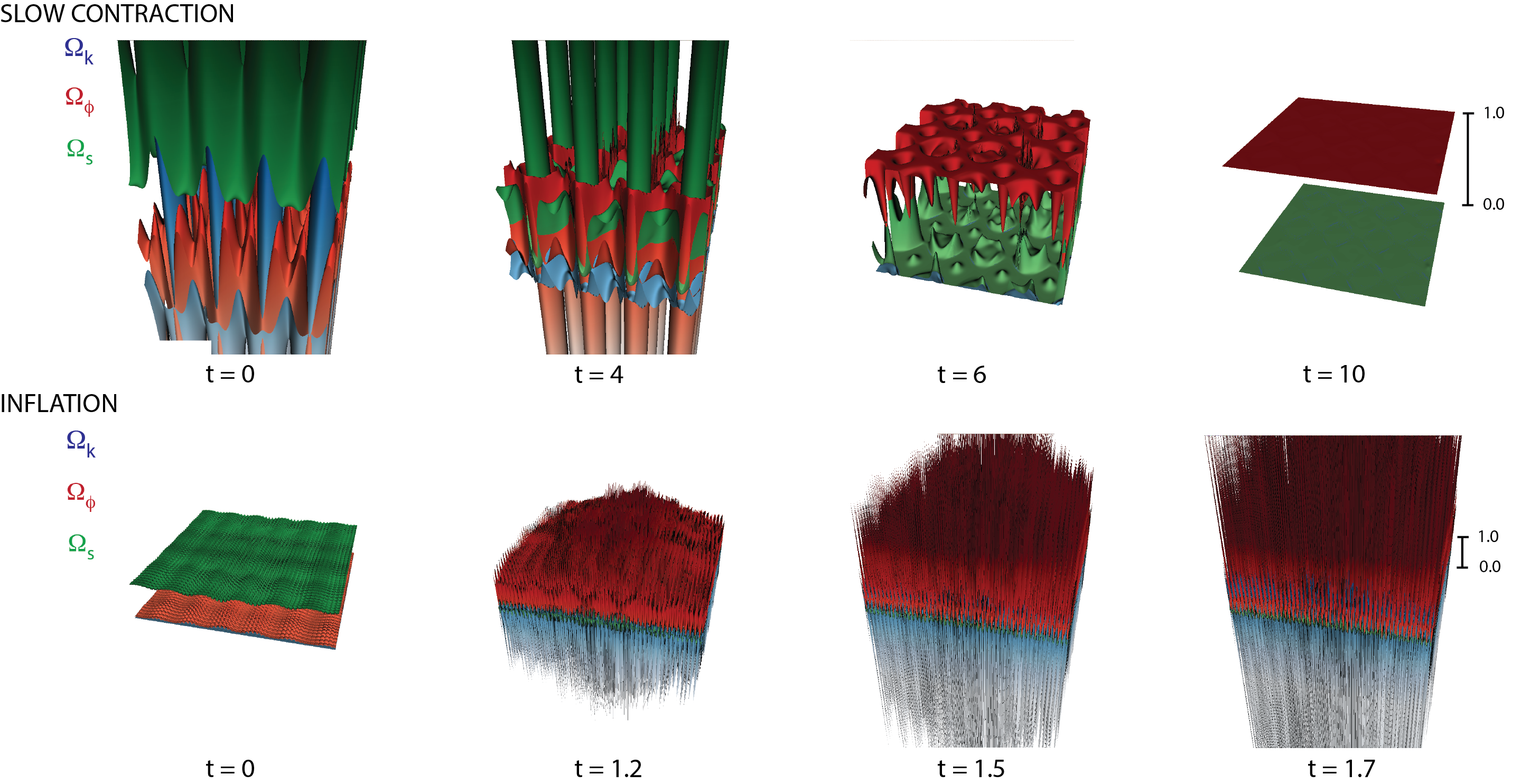}%
    \caption{Snapshots show the relative contributions of scalar field energy, shear and curvature   ($\Omega_{\phi, s, K}$) in the initial time step and then a series  of time steps following.  In the contraction case (top row), the initial spatial gradients are so large that they extend beyond the frame, but they are eventually smoothed through ultralocality and slow contraction.  In the inflation case (bottom row), the initial spatial gradients are small and grow rapidly until they extend outside the frame, consistent with the idea that the opposite of ultralocality occurs when initial conditions are far from flat FRW. 
    }   
    \label{Fig7}%
\end{figure}

Fig.~\ref{Fig7} plots  $\Omega_{\phi, s, K}$, the relative contributions of scalar field energy, shear and curvature, as a function of time for two cases.  The first is the slow contraction model described in Sec.~\ref{sec5} and Table~\ref{Fig1} that begins with conditions so extremely far from flat FRW that the plot extends  far outside the frame for the initial time steps,  However, after 15 $e$-folds of contraction, the combination of ultralocality and slow contraction smooths.  

The second case is  an inflation example where the spatial gradients begin small and then grow rapidly during the first steps.   This is the expected signature if the opposite of ultralocality occurs and is the reason why smoothing fails. 

%Figure7

\section{Discussion}
\label{sec8}

Numerical relativity can play a crucial in exploring fundamental issues in cosmology such as considered here: the solution to the homogeneity, isotropy and flatness problems.   We have used objective tests, including ones based on gauge/frame invariant measures, and found huge differences between slow contraction and inflation.  Slow contraction can smooth and flatten even when the initial conditions are exponentially far from flat FRW initial conditions.  
Inflation can only smooth in cases where the initial conditions are comparatively close to flat FRW at the start.  The simulations also enable us to understand why there is this enormous qualitative and quantitative difference.   Physics becomes ultralocal in a contracting universe which rapidly reduces spatial gradients until they become negligible for the dynamics.  When combined with slow contraction ($\varepsilon \gg 3$),   world-lines converge to a flat FRW attractor.   Conversely, spatial gradients grow in an expanding pre-inflating universe.  If they are  large to begin with, as one would expect after a big bang, they can grow fast enough that inflation cannot take hold. 

These results were indicated in some earlier studies, although other studies suggested inflation is also an effective smoother.   Important advances incorporated in this study are formulations that allow a much broader range of initial conditions for both inflation and slow contraction and  objective diagnostics that enable the distinction between initial conditions that are outside the perturbative regime of flat FRW.   Including these leads to the unambiguous conclusions presented here -- and this despite the limitations in current numerical methods described in Sec.~\ref{sec4} that favor inflation or our choice of a positive quadratic potential for inflation that has been found to be least sensitive to initial conditions
\cite{Clough:2016ymm,Aurrekoetxea:2019fhr,Clough:2017efm}.  (As noted in the introduction, we have extended our survey to other potentials and will  report the results elsewhere; they do not appear to change the qualitative conclusions presented here.)

The results have potentially profound implications.  
 Observations inform us that we are living in a spacetime that is  extraordinarily special -- much more homogeneous, isotropic and flat than one would expect from general relativity, or from a multiverse, or from what can be inferred based on anthropic arguments.  Some remarkable predictive physical process must account for it.  If the universe does not emerge perturbatively close to homogeneous and isotropic immediately after a big bang, the results suggest that adding an inflaton does not typically smooth or flatten it, thanks in part to the absence of ultralocality.  At the same time, it appears that ultralocality in a slowly contracting universe can accomplish the task in a predictive, universal and acausal way.

\subsection*{Acknowledgements.} We thank D. Shlivko for useful conversations and M. Klatt for helping to create some of the Python visualization scripts. 
The work of A.I. is supported by the
Simons Foundation grant number 947319.  P.J.S. is supported in part by the DOE grant number DEFG02-91ER40671 and by the Simons Foundation grant
number 654561. DG is supported in part by NSF grant number PHY-2102914.  
\bibliographystyle{apsrev}
\bibliography{CompContractionInflation}

\end{document}